\documentclass[10pt,letterpaper]{article}
\usepackage[top=0.85in,left=2.75in,footskip=0.75in]{geometry}
\usepackage{amsmath,amssymb,mathtools}
\usepackage{changepage}
\usepackage[utf8]{inputenc}
\usepackage{textcomp,marvosym}
\usepackage{cite}
\usepackage{nameref,hyperref}
\usepackage[right]{lineno}
\usepackage{microtype}
\usepackage{graphicx}
\usepackage{array}
\usepackage{booktabs}
\usepackage{color}
\usepackage[table]{xcolor}
\usepackage{caption}
\usepackage{enumitem}
\usepackage{algorithm}
\usepackage{algpseudocode}
\usepackage{fancyhdr}
\fancyheadoffset[L]{2.25in}
\fancyfootoffset[L]{2.25in}
\usepackage{lastpage,ifthen}

\newcommand{\Prob}{\mathbb{P}}

\newtheorem{proposition}{Proposition}

\begin{document}
\vspace*{0.2in}

\begin{flushleft}
{\Large
\textbf\newline{What should we forget?
A computational model of memory consolidation}
}
\newline
\\
Xin Li\textsuperscript{1,*}
\\
\bigskip
\textbf{1} Department of Computer Science, University at Albany, Albany, NY 12222
\\
\bigskip
* xli48@albany.edu
\\
\bigskip
\textbf{Short title:} Predictive coarse-graining of biological memory
\end{flushleft}

% ============================================================
\section*{Abstract}
Neural and immune memory rely on different biological mechanisms but face the
same computational problem: future situations rarely repeat past ones exactly.
Memory must retain distinctions that alter future responses while
discarding irrelevant variation. We formulate this problem as
\emph{scaffold-flow memory}: fast, state-dependent responses constitute the
flow, whereas slowly changing physical variables form a scaffold that constrains
future dynamics. Consolidation writes a predictive coarse-graining of experience
into that scaffold. A useful coarse-graining must preserve future-relevant
distinctions, generalize to novel experiences, support approximately autonomous
coarse dynamics, and provide enough future benefit to justify its physical cost.
We quantify failures of coarse autonomy through a leakage measure, relate leakage
to excess future error, and identify persistent memory classes with slow
dynamical modes. We further show that when experience does not self-average,
storage is necessary rather than efficient. In a Willshaw associative
memory, a metastable neural attractor, and a stochastic affinity-maturation
model, future risk is minimized at an intermediate granularity: overly fine
representations waste capacity and generalize poorly, whereas overly coarse ones
merge situations requiring different responses. These results support a common computational
principle: \emph{memory consolidation selects a predictive, dynamically usable, and
affordable representation of the past}.

% ============================================================
\section*{Author summary}
Because no experience repeats exactly, memory cannot operate by storing exact
copies and waiting for them to recur. It must learn which differences matter for
future behaviour. We propose that this is a shared computational problem for
brains and adaptive immune systems despite their different biological machinery.
In both, fast responses are shaped by slower physical structure: synaptic and
cellular properties in neural circuits, and clone abundance, receptor properties,
and differentiation state in immunity. Consolidation determines which
experiences should share a persistent category and writes that category into the
slow structure. A useful category must preserve consequential differences,
include novel variants, remain usable without recovering discarded microscopic
detail, and justify its biological cost. Simulations of three distinct systems
show the same trade-off: categories that are too narrow consume resources and
fail to generalize, whereas categories that are too broad cause false recall or
self-reactivity. The preferred granularity lies between these extremes and shifts
with environmental variability, memory load, and the cost of false alarms.

%\linenumbers

% ============================================================
\section*{Introduction}

Biological systems encounter streams of situations that never repeat exactly.
They must decide which differences are worth preserving. This is a
compression problem in which the target is not the original experience but its
future consequence. Ergodicity closes the problem from the opposite direction
\cite{walters2000introduction}: if experience were stationary and rapidly
mixing, omitted regularities could eventually be recovered by observing longer.
Behaviourally important events, however, may occur once, mix more slowly than an
organism's lifetime, or arise from processes whose time averages do not settle.
Such structure cannot simply be recomputed later. Because storing everything is
also impossible, memory must retain a grouping that supports generalization
\cite{bousquet2002stability}.

\begin{figure}[!ht]
\centering
\includegraphics[width=\textwidth]{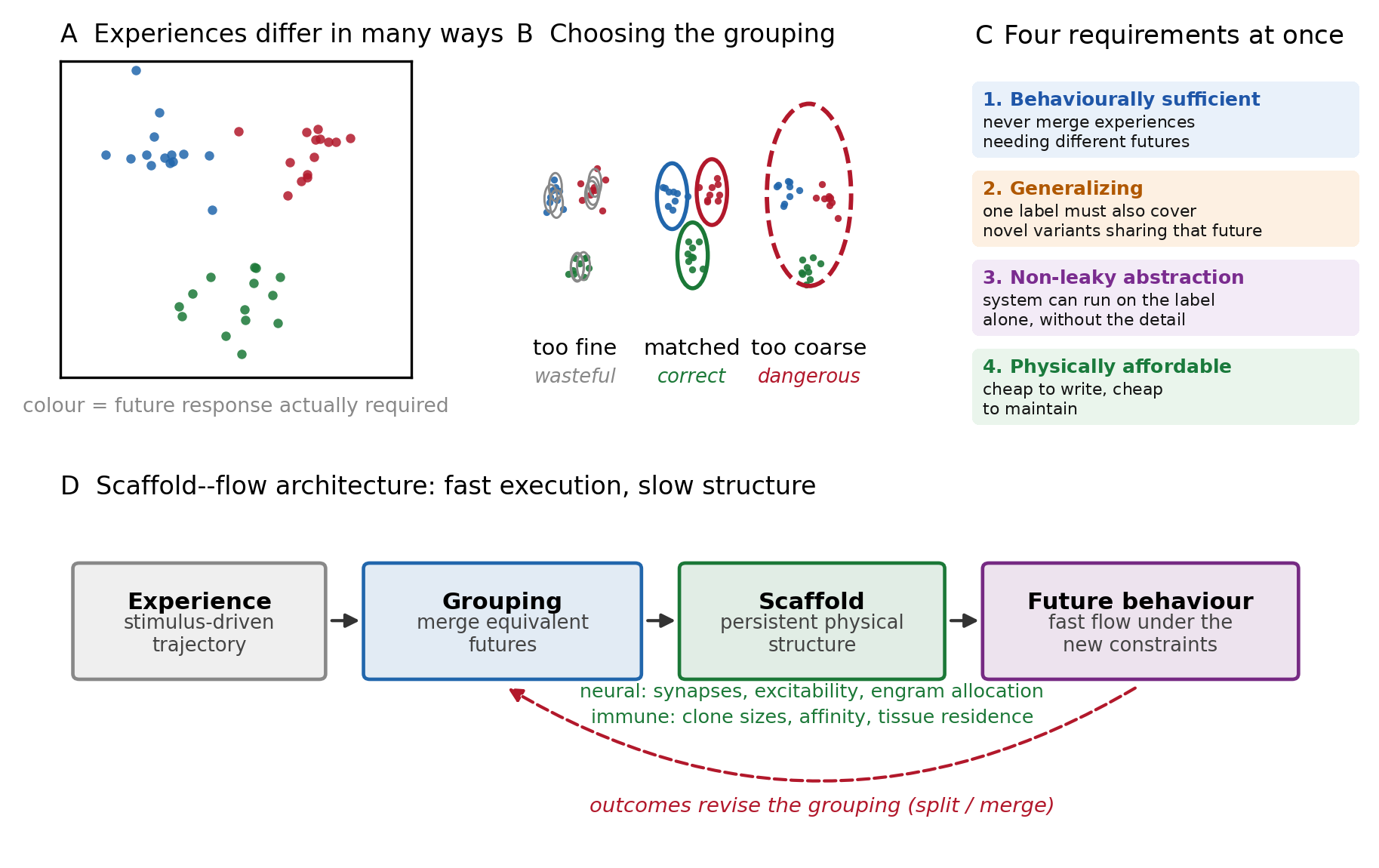}
\caption{{\bf The computational problem and the proposed architecture.}
(A) Experiences differ along many dimensions; only some differences change what
should happen next (colour). (B) Grouping too finely wastes capacity and
generalizes poorly to novel variants; grouping too coarsely merges experiences
requiring different responses. (C) The four requirements a useful grouping must
satisfy simultaneously. (D) The scaffold-flow loop: experience is grouped, the
grouping is written into slow physical structure, and that structure constrains
future fast dynamics; outcomes feed back to split or merge groups. The scaffold
is realized in unrelated variables in the two biological systems.}
\label{fig:schematic}
\end{figure}

Neural and adaptive immune systems face this problem through unrelated physical
mechanisms. Neural circuits convert episodes into synaptic and intrinsic changes
that reshape stable activity patterns \cite{Hopfield1982,Amit1989,McClelland1995}.
Adaptive immunity converts antigen encounters into altered clone abundance,
receptor affinity, differentiation state, and tissue localization
\cite{Perelson1997,DeBoer2013,Farber2014}. Yet both systems must generalize to
unseen variants, pay to maintain what they retain, and avoid costly false alarms,
such as overgeneralized fear or autoimmunity.
We formalize this shared problem as \emph{scaffold-flow memory}
(Fig.~\ref{fig:schematic}). The \emph{flow} is the fast response; the
\emph{scaffold} is the slower physical structure that shapes that response; and
consolidation writes a grouping of experiences into the scaffold. A useful
grouping must satisfy four requirements simultaneously:

\begin{enumerate}[leftmargin=1.6em,itemsep=1pt]
\item \textbf{Behavioural sufficiency.} Experiences may be merged only when
they induce the same relevant distribution over futures.
\item \textbf{Generalization.} The grouping must assign novel experiences by
means of an address that includes unseen variants with the same future while
excluding variants that require a different response.
\item \textbf{Non-leaky abstraction.} Once an experience is labelled, its
coarse dynamics must be predictable from the label alone; otherwise the
abstraction still depends on discarded microscopic detail.
\item \textbf{Affordability.} The grouping must be writable and maintainable
in physical structure at acceptable cost.
\end{enumerate}

These requirements connect established formalisms but are rarely imposed
jointly. Causal states and input-output $\epsilon$-transducers formalize
predictive sufficiency \cite{CrutchfieldYoung1989,Shalizi2001,BarnettCrutchfield2015};
predictive information bottlenecks and rate--distortion extend the idea to
finite resources and finite horizons \cite{TishbyPereiraBialek1999,Still2014}.
Markov lumpability \cite{KemenySnell1976,Buchholz1994}, probabilistic bisimulation \cite{LarsenSkou1991}, and bisimulation metrics \cite{Ferns2004,Ferns2011}
characterize when coarse dynamics are autonomous and when similarity licenses
extrapolation to unseen states.
Thermodynamics supplies a complementary account of implementation cost: retaining
non-predictive information can contribute to dissipation, although our objective
is not derived from a specific thermodynamic model
\cite{StillSivakBellCrooks2012}.

The joint demand is substantive. A grouping can preserve a designated future
statistic yet fail to support autonomous coarse dynamics; it can fit observed
episodes yet leave novel variants unassigned; or it can be predictive and usable
but too costly to maintain. We treat consolidation as constrained model
selection over both coarse-grainings and their physical realizations. This view
also casts scaffold-flow memory as adaptive hysteresis
\cite{mayergoyz2003mathematical}: different experiential paths can leave
different persistent scaffold states, causing the same present cue to generate
different subsequent dynamics.

% ============================================================
\section*{Results}

\subsection*{Memory consolidation as a computational algorithm}

We first state the proposal operationally. The memory system maintains groups
with an address $a_g$ specifying which experiences recruit them, a response
program $r_g$, and running future statistics $F_g$. For each experience, the
system recalls the best-matching group and updates $F_g$; splits a group whose
members require different futures; broadens or contracts its address according
to whether new variants share that future; merges groups with equivalent
futures; and writes the resulting addresses and responses into the scaffold.
Scaffold updates reduce dynamical leakage, while unused groups decay. The full
procedure appears as Algorithm~\ref{alg:consolidate} in \nameref{sec:mm}.

\paragraph{What distinguishes scaffold-flow consolidation?}
First, split and merge decisions are made in \emph{predictive space}, not from
surface similarity alone. Addresses propose candidate matches, but group
membership is revised according to the future laws summarized by $F_g$
\cite{CrutchfieldYoung1989,Shalizi2001,BarnettCrutchfield2015,Still2014}.
Superficially similar episodes must separate when their consequences
differ, whereas dissimilar episodes may merge when those consequences agree.
With finite data, these are distortion-bounded clusters rather than exact
equivalence classes.
Second, persistence is a continuing resource-allocation decision. Recruitment
refreshes a group's statistics, address, and response, whereas prolonged nonuse
allows the trace to weaken, consistent with active forgetting and immune
turnover \cite{Hardt2013,DeBoer2013,Farber2014}. Frequency supplies evidence of
future value, but retention is ultimately justified by expected risk reduction
relative to writing and maintenance cost.
Third, consolidation changes not only the partition but also the dynamics that
realize it. Conventional clustering updates assignments while leaving the data-
generating dynamics untouched \cite{duda2006pattern,MacQueen1967,Aggarwal2003}.
A memory abstraction must additionally make states within each group dynamically
interchangeable at the coarse level, as required by lumpability \cite{KemenySnell1976,Buchholz1994} or probabilistic
bisimulation \cite{LarsenSkou1991,Ferns2004,Ferns2011}.

\paragraph{Leakage: why behavioural sufficiency is not enough}
A predictively sensible grouping need not support autonomous dynamics. Suppose
states A and B are assigned to one group because they share an immediate future,
but they reach the next group at different rates. The group label then fails to
predict its own evolution even though the designated immediate statistic was
preserved.
We quantify this failure by the \emph{leakage} of a grouping $\pi$ over horizon
$T$,
\begin{equation}\label{eq:leak}
\Lambda_T(\pi)=\max_{\pi(x)=\pi(x')}
\ \text{distance}\Big(\text{future of }x,\ \text{future of }x'\Big),
\end{equation}
where futures are compared after projection onto group labels (precise definition
in \nameref{sec:mm}). A system that acts on labels rather than microscopic
states incurs excess future error bounded by
\begin{equation}\label{eq:excess}
\text{excess error}\ \le\ L\,\Lambda_T(\pi)\ +\ \bar\ell\,\Prob(\text{escape from group before }T),
\end{equation}
where $L$ is the sensitivity of loss to changes in the future distribution and
$\bar\ell$ is the worst-case loss. Because
leakage depends on both the grouping and the generator, consolidation may either
refine the partition or alter the scaffold so that the intended partition
becomes non-leaky. This joint adaptation distinguishes memory formation from
ordinary categorization.

\subsection*{Consolidated neural categories appear as slow modes}

\begin{figure}[!ht]
\centering
\includegraphics[width=\textwidth]{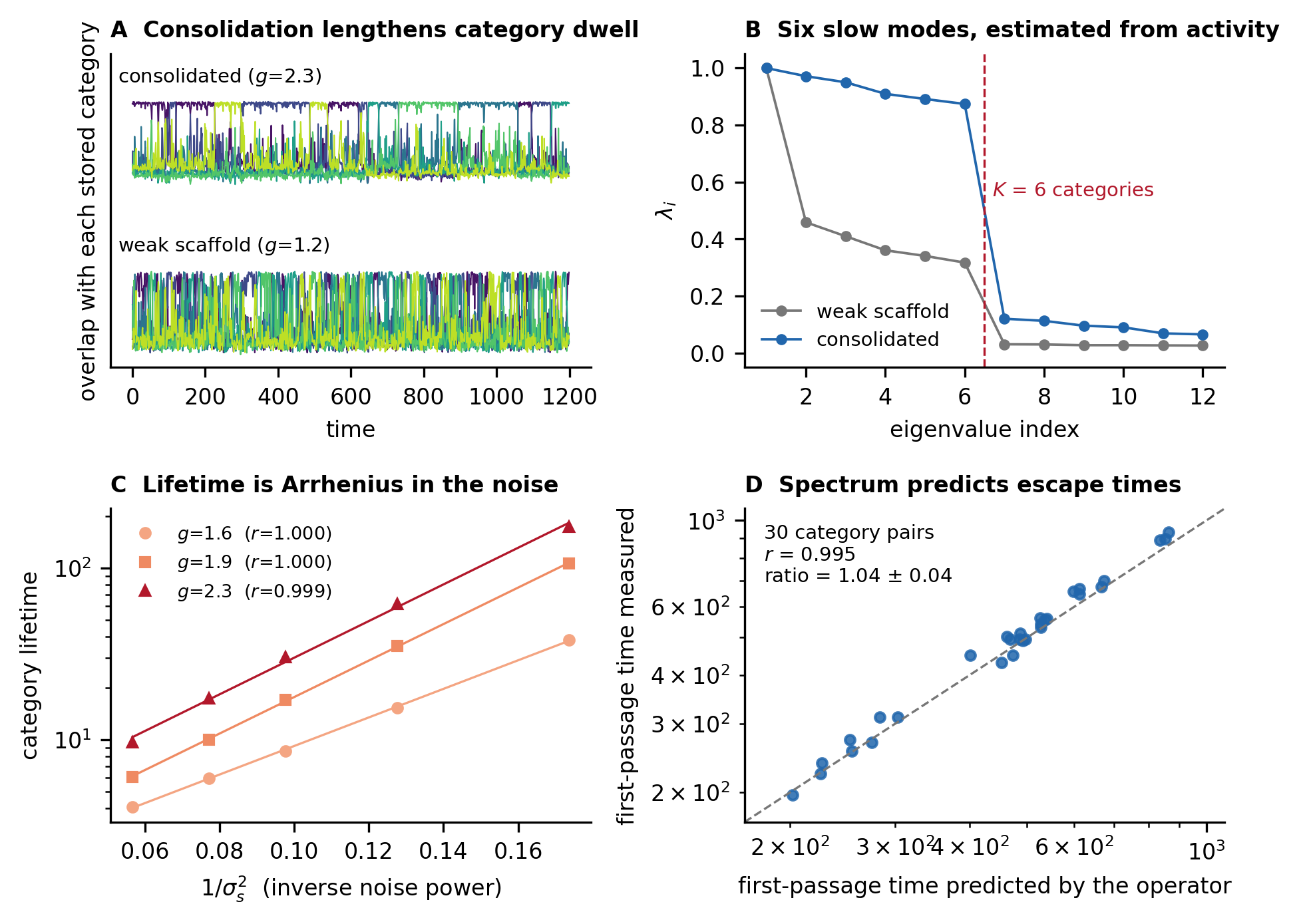}
\caption{{\bf Consolidation converts fast fluctuations into persistent
categories in a recurrent network.}
$N=500$ rate units store $K=6$ sparse patterns; consolidation raises the
recurrent gain $g$. Panels (B) and (D) use only the activity of a random half of
the units and never the stored patterns.
(A) Overlap with each stored pattern over time. With a weak scaffold ($g=1.2$)
the network wanders among the six categories; after consolidation ($g=2.3$) it
dwells in each, mean dwell rising from $4.17$ to $48.02$.
(B) Eigenvalues of the transition operator estimated from the activity: six
modes are separated from the rest in both conditions, but consolidation moves
them from $\lambda\approx0.3$-$0.46$ to $0.87$-$0.97$, i.e.\ converts
fluctuations into structure. Macrostate purity against the stored patterns is
$98.9\%$ after consolidation.
(C) Category lifetime is Arrhenius in the shared-noise power, at three scaffold
depths ($r\ge0.9987$ each); the fitted slope is the barrier height.
(D) First-passage time predicted by the operator versus directly measured, for
all $30$ ordered category pairs ($r=0.9952$, ratio $1.04\pm0.04$). The dashed
line is the identity.}
\label{fig:neural}
\end{figure}

If consolidation makes memory categories self-contained and persistent, they
should appear in the system's dynamics as long-lived regions separated by rare
transitions. Transfer-operator theory gives a direct signature: near-unit
eigenvalues enumerate metastable categories, and their gaps from unity determine
the associated transition timescales
\cite{DellnitzJunge1999,SchuetteSarich,BovierGayrardKlein}.
We tested this prediction in a recurrent network of $N=500$ rate units storing
$K=6$ sparse patterns. Consolidation increased the recurrent gain $g$, thereby
deepening basins without changing their identities (Fig.~\ref{fig:neural}A).
To preserve observable collective transitions at this size, the network was
driven by structured variability along the coding directions. All spectral
analyses used only the activity of a randomly selected half of the units and had
no access to the stored patterns: recorded states were clustered into
microstates, a transition operator was estimated at lag $5$, and macrostates
were recovered from its leading eigenvectors.

The four panels form a quantitative chain from scaffold strength to memory
persistence. Mean dwell time rose from $4.17$ to $48.02$ as $g$ increased from
$1.2$ to $2.3$, while overlap with the occupied pattern rose from $0.798$ to
$0.958$ (Fig.~\ref{fig:neural}A). The estimated spectrum contained exactly six
slow modes, followed by a sharp drop; consolidation shifted these modes toward
unity and increased the longest implied timescale from $6.4$ to $172.7$ without
changing their number (Fig.~\ref{fig:neural}B). The recovered macrostates matched
the stored categories with $98.9\%$ frame-wise purity. Category lifetime also
obeyed Arrhenius--Kramers scaling \cite{kramers1940brownian}: at each scaffold depth, log-lifetime was linear
in $1/\sigma_s^2$ with $r=0.9987$--$0.9999$, and the fitted barrier increased
from $19.1$ to $24.6$ (Fig.~\ref{fig:neural}C). Finally, first-passage times
inferred from the operator predicted directly measured values across all $30$
ordered category pairs at $r=0.9952$, with ratio $1.038\pm0.044$
(Fig.~\ref{fig:neural}D). Memory categories are measurable as slow
modes rather than described by analogy.

\begin{figure}[!ht]
\centering
\includegraphics[width=\textwidth]{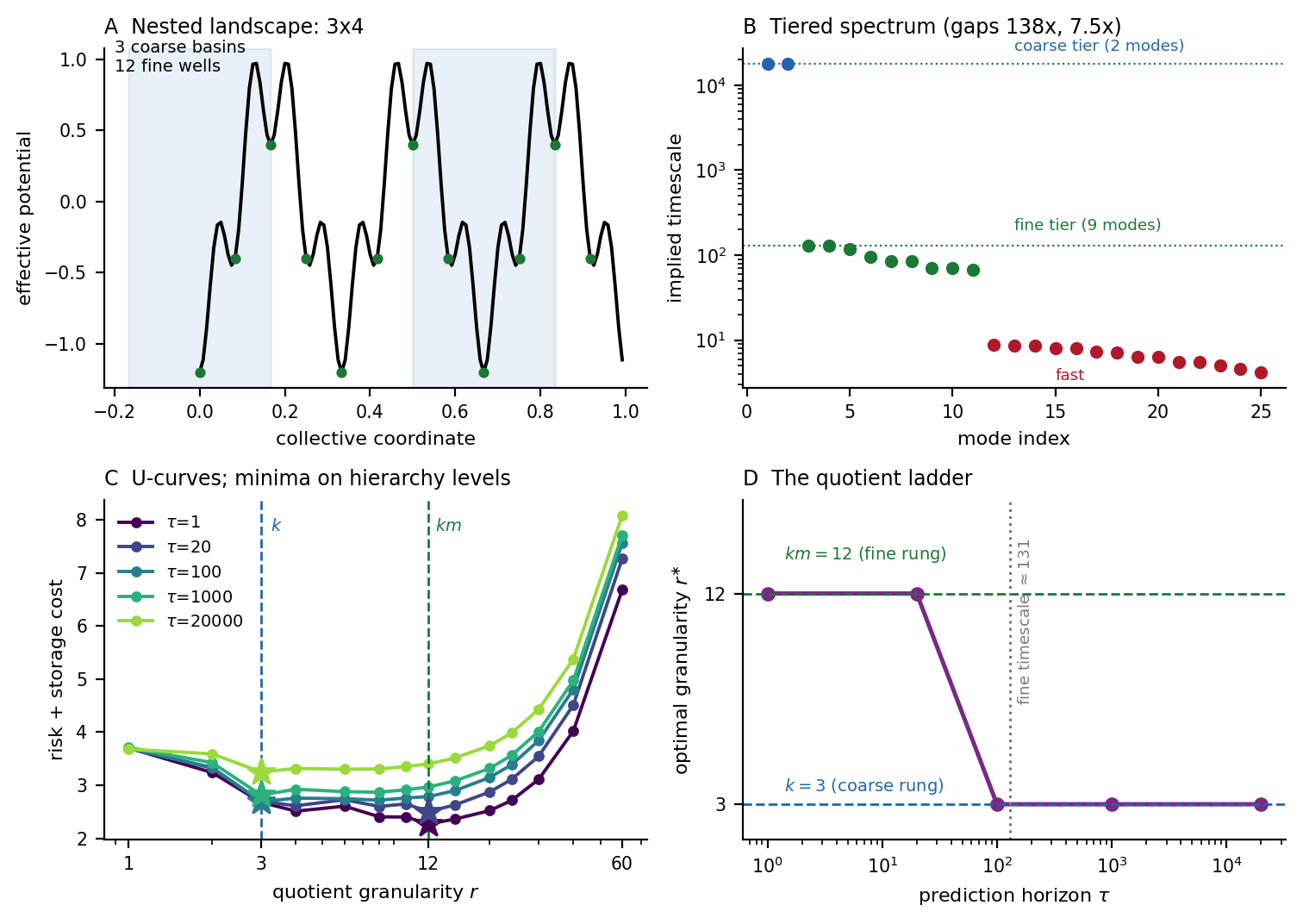}
\caption{{\bf Optimal granularity lands on levels of a hierarchy, and climbs a
rung as the prediction horizon grows.}
(A) The nested landscape: $k=3$ coarse basins (shaded) each holding $m=4$
sub-basins (the true hierarchy has rungs at $r=3,12$).
(B) Implied timescales estimated from the dynamics: two coarse modes, nine fine
modes, and a fast continuum, separated by gaps of $138\times$ and $7.5\times$.
%The counts are exactly $k-1$ and $k(m-1)$.
(C) Predictive risk plus storage cost against granularity, for horizons $\tau$
from $1$ to $2\times10^{4}$. Stars mark the minima, which fall on the hierarchy
levels $r=km=12$ and $r=k=3$ (dashed lines) and never between them.
(D) The optimum as a function of horizon: it stays on the fine rung until $\tau$
crosses the fine relaxation time (dotted line), then steps to the coarse rung.
12 seeds; $3000$ transition pairs per estimate at every horizon.}
\label{fig:hier}
\end{figure}

\paragraph{Abstraction is recursive: the quotient has rungs}
Scaffold-flow abstraction can recur across timescales
\cite{hasson2015hierarchical}: a completed coarse-graining may itself become the
scaffold for a slower one. We tested this idea in a ring landscape with a known
hierarchy of $k=3$ coarse basins, each containing $m=4$ sub-basins. The spectrum
contained two coarse modes with implied timescale approximately
$1.8\times10^4$, nine fine modes at $67$-$131$, and fast modes near $8$, with
clear gaps between tiers (Fig.~\ref{fig:hier}B).
For each candidate granularity $r$, we estimated the out-of-sample
negative log-likelihood of the future microstate using an equal budget of $3000$
transition pairs. Without a capacity penalty, the fine level $r=12$ was optimal
for horizons $20$--$500$. Adding the description-length cost of the coarse
transition model restricted the optimum to the two true hierarchy levels: it
lay at $r=12$ for $\tau\le50$ and moved to $r=3$ from $\tau=100$ onward, as
the horizon crossed the fine relaxation time (Fig.~\ref{fig:hier}C,D). The
$r=12$ quotient retained both the fine and coarse timescales, whereas the $r=3$
quotient retained only the coarse one. Coarse-graining at one rung
discards faster detail while preserving slower structure, allowing the resulting
quotient to serve as a scaffold at the next level.

\subsection*{In associative memory, discarding detail pays only under load and asymmetric costs}

\begin{figure}[!ht]
\centering
\includegraphics[width=\textwidth]{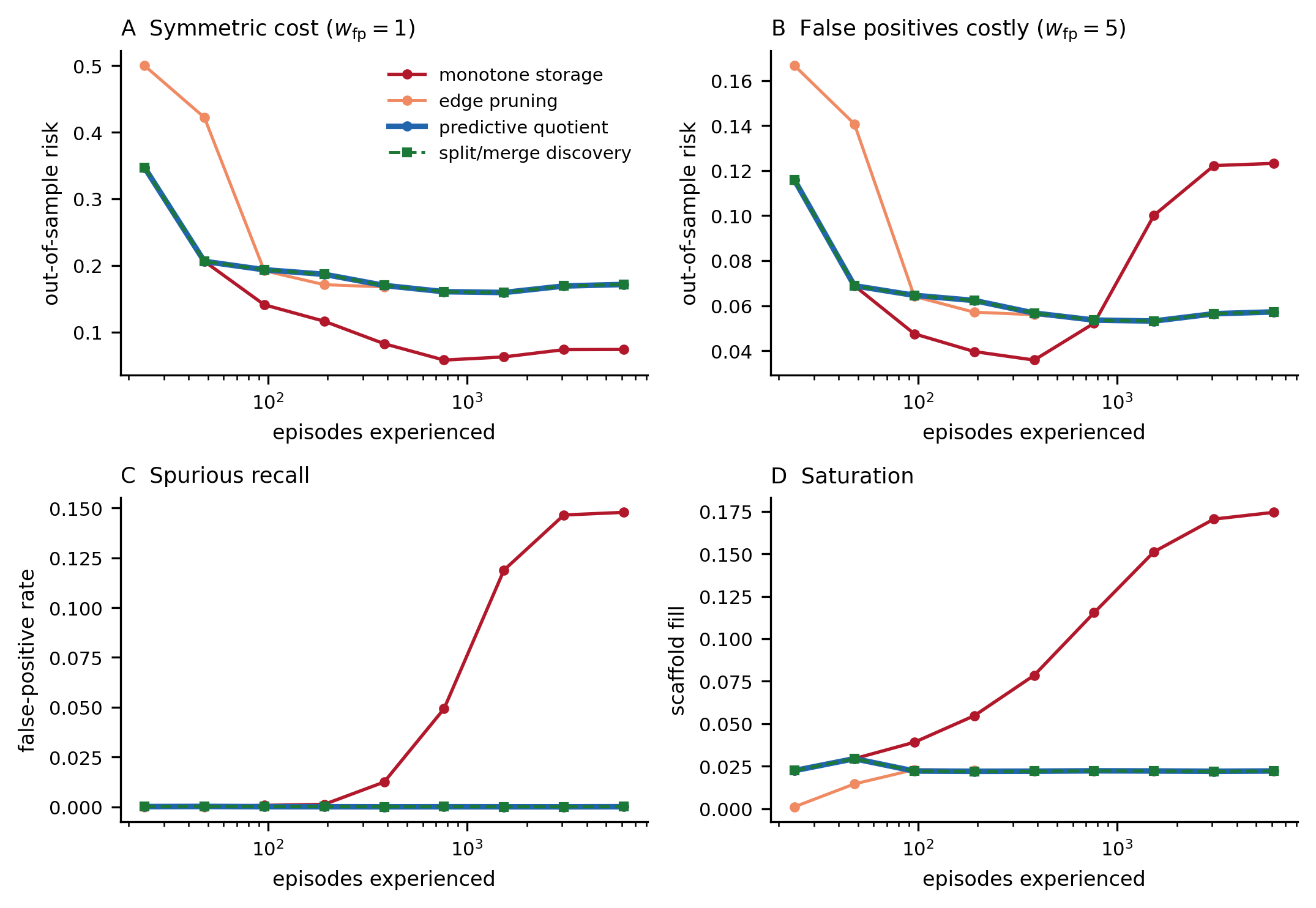}
\caption{{\bf Grouping beats storing under load when false alarms are costly.}
(A) With false alarms and misses weighted equally, storing every experience wins
at all loads. (B) With false alarms weighted five-fold, storage degrades sharply
beyond $\sim10^3$ experiences and grouping wins. (C) Storage accumulates spurious
recall; grouping does not. (D) Grouping saturates at a load-independent level of
structure. Split/merge discovery (dashed) coincides with the oracle grouping.
Mean of 6 runs.}
\label{fig:willshaw}
\end{figure}

We next asked when grouping outperforms episodic storage, using a Willshaw
associative memory \cite{WillshawBuneman1969}. The task contained six predictive
response categories, each represented by four cue clusters and many corrupted
experiences. We compared storing every experience, count-based edge pruning,
oracle grouping at the predictive level, and split/merge discovery.
Episodic storage accumulated corruption as spurious edges: at the highest load,
its false-alarm rate reached $0.148$ and it occupied $17.5\%$ of the matrix,
whereas grouping held false alarms near $10^{-4}$ with $87.3\%$ less structure
(Fig.~\ref{fig:willshaw}C,D). Yet with equal costs for false alarms and misses,
storage remained better because grouping also discarded useful detail
(Fig.~\ref{fig:willshaw}A). When false alarms were weighted five-fold, grouping
overtook storage at roughly $1500$ experiences and ended at error $0.057$ versus
$0.123$ (Fig.~\ref{fig:willshaw}B). Therefore, compression is advantageous not
universally, but under sufficient load and asymmetric error costs. Split/merge
discovery came within $0.02$ error of the oracle without being given the category
structure, showing that predictive equivalence can be learned from outcomes.

\paragraph{A quotient needs the cue and the future stored separately}
The Willshaw experiment reveals when grouping pays but not where the quotient
should act. Predictive grouping must merge experiences on the side of the
\emph{future} while preserving fine cue addresses. Classical and dense
Hopfield memories remain autoassociative, so their keys and values cannot be
coarsened independently \cite{Hopfield1982,krotov2016dense}. Modern Hopfield
retrieval, $\hat v=V^{\top}\mathrm{softmax}(\beta Kq)$, separates keys and
values and can express this asymmetry \cite{ramsauer2020hopfield}.
We compared episodic storage, metric clustering in cue space, an oracle
predictive quotient with $24$ fine keys and $6$ pooled values, and split/merge
discovery of those values.

\begin{figure}[!ht]
\centering
\includegraphics[width=\textwidth]{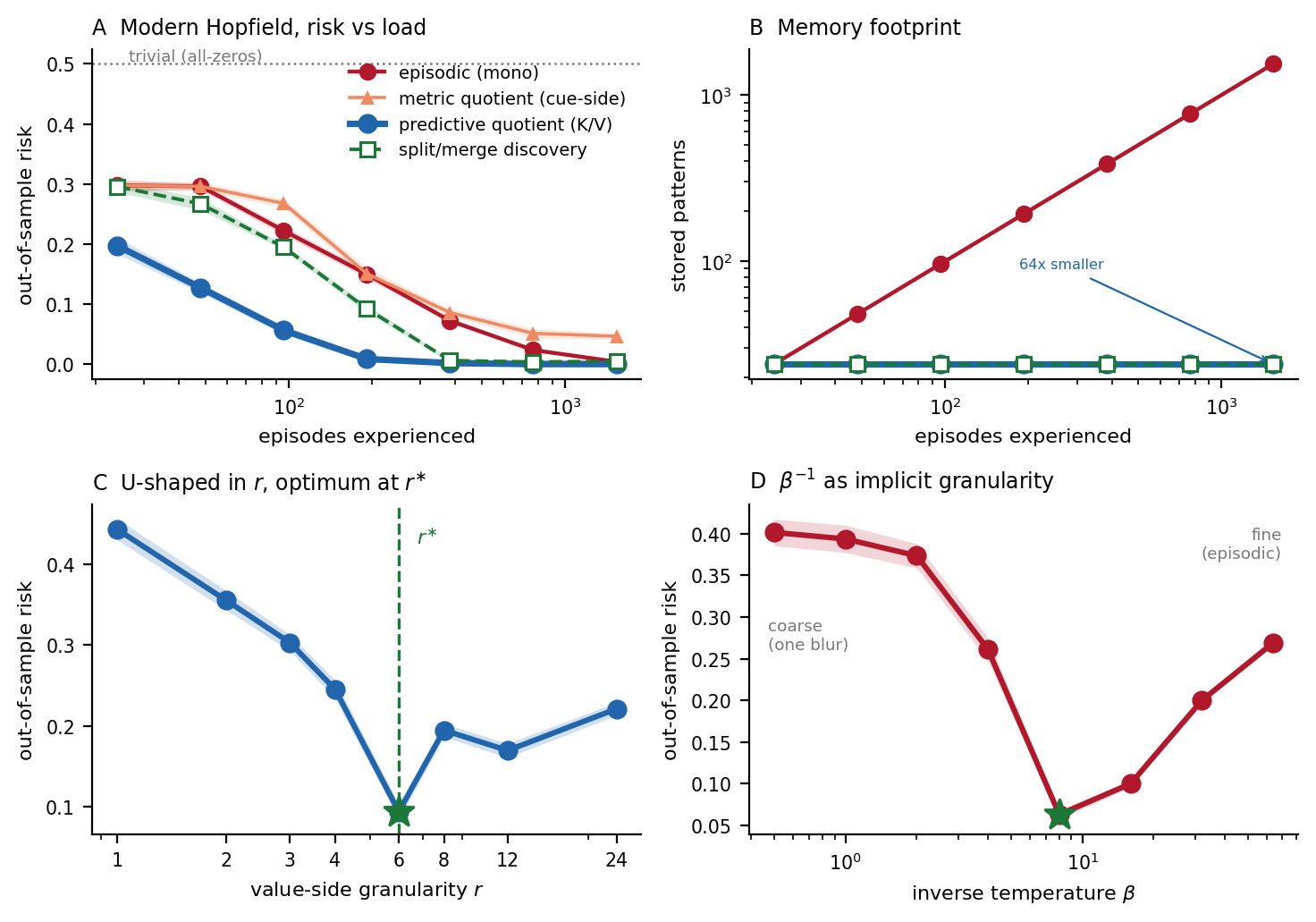}
\caption{{\bf In a modern Hopfield memory the quotient acts on the value side, and
the retrieval temperature is a granularity knob in its own right.}
(A) Out-of-sample risk against load for four strategies. The predictive quotient
(fine keys, $6$ pooled values) dominates at every load; split/merge discovery
converges to it once each cue cluster has been seen about $16$ times.
(B) Stored patterns: at the highest load the quotient holds $24$ against episodic
storage's $1536$.
(C) Sweeping the value-side granularity $r$ gives a U-curve with its minimum at
$r^*=6$. The point at $r=8$ is off-trend because $8$ does not divide the cue
structure, so its groups straddle predictive boundaries.
(D) With no explicit grouping at all, risk is U-shaped in the inverse temperature
$\beta$, because $\beta$ sets how far retrieval averages over similar stored
patterns. Eight seeds; bands are $\pm1$ s.e.m.}
\label{fig:hopfield}
\end{figure}

The predictive quotient dominated at every load and reached zero out-of-sample
risk (Fig.~\ref{fig:hopfield}A). At the highest load it matched or exceeded
episodic storage while retaining $24$ patterns instead of $1536$
(Fig.~\ref{fig:hopfield}B). Split/merge discovery converged to the oracle after
roughly $16$ episodes per cue cluster. A sweep over value-side granularity gave
the expected U-shaped risk with its minimum at the true number of futures,
$r^*=6$ (Fig.~\ref{fig:hopfield}C).
The retrieval dynamics also provide an implicit granularity control. At low
inverse temperature $\beta$, attention averages over too many memories; at high
$\beta$, it retrieves noisy episodes individually. With no explicit grouping,
risk was U-shaped in $\beta$ and minimized at $\beta=8$
(Fig.~\ref{fig:hopfield}D). A metastable average over similar values functions
as a quotient, showing how consolidation can tune granularity through a physical
parameter of the retrieval dynamics rather than through an external data
structure.

\subsection*{In immunity, grouping coarseness emerges from selection}
 
The immune realization uses a different set of scaffold variables, including
clone locations in antigenic shape space, receptor breadth, and clonal
abundance \cite{perelson1979theoretical}. Unlike the associative-memory
experiment, the grouping is not prescribed in advance but emerges through
mutation, competition, and affinity-dependent clonal selection. We model
germinal-centre affinity maturation as a stochastic clonal-selection process,
reflecting the repeated diversification and selection of B cells and the role
of follicular helper T-cell signals in preferentially expanding successful
clones \cite{victora2012germinal,gitlin2014clonal}.

Each simulated B cell carries a receptor position and a breadth parameter. We
impose a stylized affinity-breadth trade-off motivated by observations that
germline or early-memory antibodies can exhibit low-affinity cross-reactivity,
whereas somatic maturation often increases affinity and specificity, although
broad reactivity can also emerge through affinity maturation toward conserved
epitopes \cite{yin2003structural,takahashi2017role}. Thus, the trade-off used
here is a modeling abstraction rather than a universal law of antibody
evolution. Pathogens are antigenically variable: each of six simulated
families presents several strains. A single selection exponent $p$ controls how
the model aggregates a receptor's affinities $A_i$ across the strains of its
family through the power mean
$\left(\frac{1}{n}\sum_i A_i^p\right)^{1/p}$.
Large positive $p$ favors clones with a strong match to at least one strain,
$p\to-\infty$ favors clones with strong worst-case coverage, and intermediate
values interpolate between these selection objectives. Receptor breadth is not
assigned to a predetermined clone class but evolves under mutation and this
selection rule.

The simulation also imposes a finite repertoire budget, motivated by
homeostatic competition among B-cell populations for limited survival
resources \cite{mclean1997resource}. Specifically, the matured repertoire is
restricted to $240$ cells regardless of how finely it is partitioned, and at
least $K=16$ precursors must bind before a variant is counted as neutralized
or a self target as autoreactive \cite{vinuesa2009dysregulation}. These numerical values are simulation
parameters. Their purpose is to make over-refinement costly: splitting a fixed
repertoire among more groups leaves fewer cells supporting each group and can
therefore reduce reliable recall against antigenically drifted variants.
 
\begin{figure}[!ht]
\centering
\includegraphics[width=\textwidth]{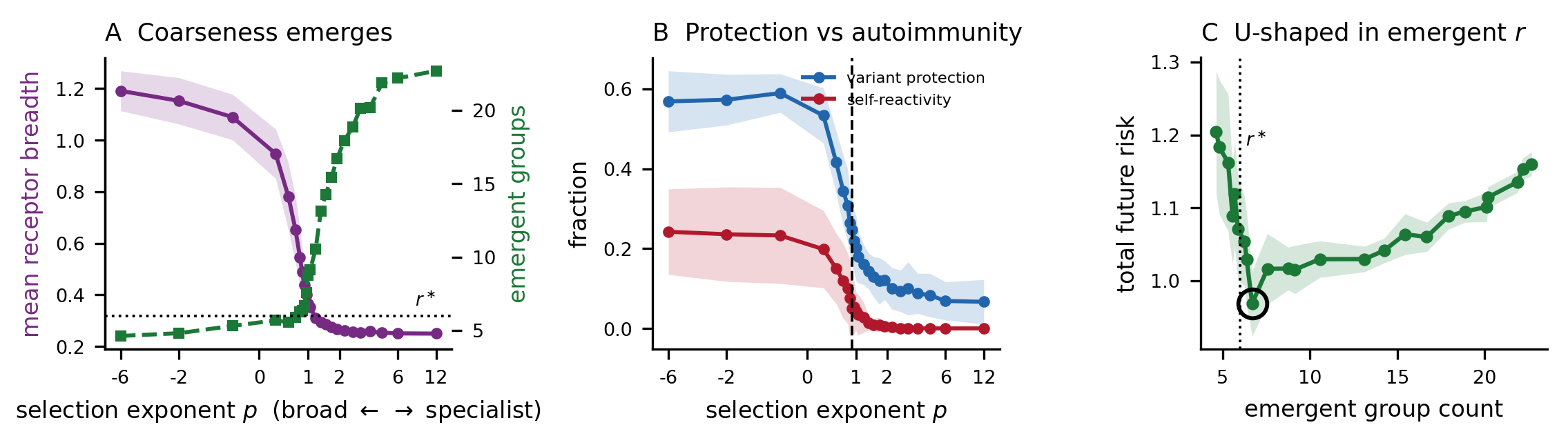}
\caption{{\bf In affinity maturation, grouping coarseness emerges from selection
pressure.}
(A) As the selection exponent $p$ falls, generalization pressure rising from
right to left, mean receptor breadth increases (purple) and the number of
emergent memory groups falls (green): a coarser quotient, produced by selection
rather than imposed. The horizontal axis is $\mathrm{arcsinh}\,p$, which is
linear near zero and logarithmic in the tails. (B) Broad repertoires protect
against drifted variants (blue) but past a threshold become self-reactive (red);
dashed line marks the risk optimum. (C) Total future risk versus emergent group
count is U-shaped with an interior optimum (circled) at $6.7$ groups, close to
$r^*=6$: neither the finest specialist ($22.7$ groups) nor the coarsest
generalist ($4.6$ groups) repertoire is best, and the two arms rise by
comparable amounts. Mean of 16 stochastic runs; shaded bands are $\pm1$ s.d.\ in
(A) and (B) and $\pm1$ s.e.m.\ in (C).}
\label{fig:immune}
\end{figure}

The result is that quotient coarseness is an \emph{output}
(Fig~\ref{fig:immune}). As $p$ falls, mean receptor breadth rises from
$0.250$ to $1.191$, and the number of distinct memory groups the repertoire
condenses the antigen world into measured (not imposed) falls from $22.7$
to $4.6$, bracketing the six true families. Protection against drifted variants
climbs from $0.067$ to $0.569$; but beyond a threshold, self-reactivity switches
on, rising from zero to $0.242$ as broad clones spill onto self antigens that
imperfect central tolerance left behind. Total future risk is U-shaped with an
interior optimum ($0.969$ at an emergent granularity of $6.7$ groups, versus
$1.160$ for the finest specialist repertoire and $1.204$ for the coarsest
generalist one), and the two arms are of comparable height (the over-refined arm
rises $0.191$ above the optimum, the over-merged arm $0.235$). The
two failure modes are mechanistically distinct: over-refinement produces escape
by variants, over-merging produces autoimmunity. Within the framework, they are
the same error as a mis-set split/merge threshold but now expressed in the
direction of receptor breadth. We stress this is a hypothesis the framework
generates about where the trade-off lies, not a demonstrated mechanism of
autoimmunity.
 
\subsection*{One law across three substrates}
 
\begin{figure}[!ht]
\centering
\includegraphics[width=\textwidth]{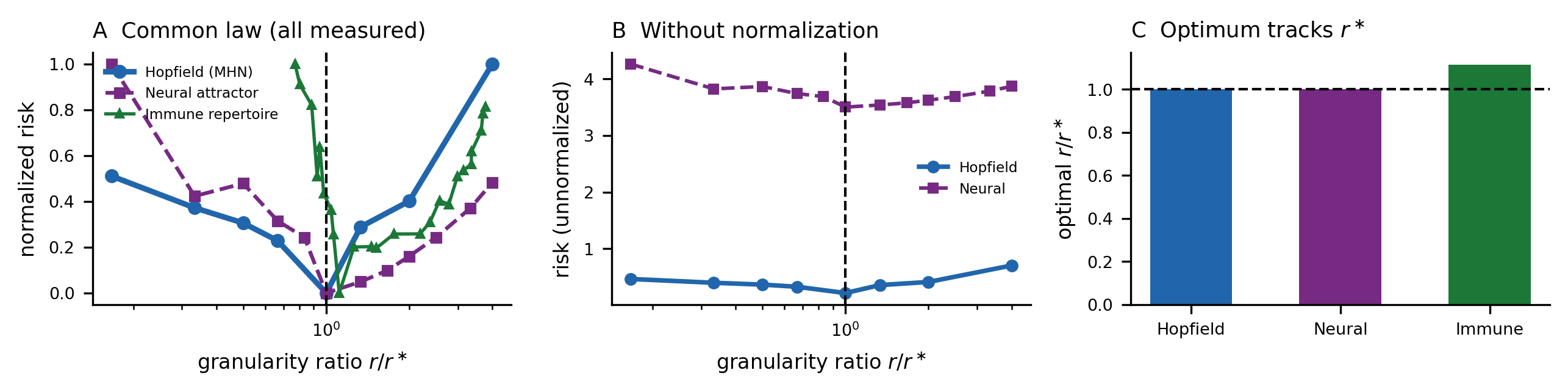}
\caption{{\bf A shared U-shaped law in three unrelated systems.}
All three curves are measured; the neural curve is the recurrent
network of Fig~\ref{fig:neural} analysed at varying grouping granularity, and the
associative curve is the value-side sweep of Fig~\ref{fig:hopfield}C with a
per-group maintenance cost added.
(A) Future error against granularity, each system min-max normalized to its own
range so the shapes can be compared; the dashed line marks $r=r^*$. Each system
has an interior optimum, but the arms are not equally steep in all three.
(B) The same associative and neural curves without normalization, showing the
raw scales the optima sit on.
(C) Optimum location in units of the true number of distinct futures $r^*$: the
two systems with a directly controlled partition sit at $r^*$, while the immune
optimum, which emerges from selection, lands at
$1.11\,r^*$.}
\label{fig:cross}
\end{figure}

The framework's sharpest cross-system prediction is that future error should be
U-shaped over the tested range of grouping granularity, with an interior optimum
rather than at either extreme. We tested this by sweeping grouping granularity
$r$ in all three systems (Fig~\ref{fig:cross}).
All three have an interior optimum, though the two arms are not equally steep in
every substrate. In the two systems where we control the partition directly, the
associative matrix and the recurrent network, the minimum occurs at the number
of genuinely distinct futures $r^*$. The associative curve here is the value-side
granularity sweep of the modern Hopfield memory of Fig~\ref{fig:hopfield}C with an
explicit per-group maintenance cost added, rather than the binary Willshaw matrix
of Fig~\ref{fig:willshaw}, because in that architecture granularity can be swept
at fixed capacity. It reaches its
lowest future error, $0.214$, at $r=r^*$, rising to $0.701$ at $r=4r^*$ and to
$0.463$ at the coarsest grouping tested; the network's predictive log-loss is
likewise minimal at $r=r^*$ ($3.503$ nats), rising to $3.870$ at $r=4r^*$ and
$4.264$ when all states are merged. The asymmetry runs in opposite directions in
the two: over-splitting is the more expensive error for the associative matrix,
over-merging for the network, so the shallow arm in one system is the steep arm
in the other. In the immune system,
where granularity is not controlled but \emph{emerges} from selection, the
optimum lands at $1.11\,r^*$ --- slightly above the number of pathogen families,
since covering each variable family well requires a little sub-family
resolution. All three optima fall within $15\%$ of $r^*$, even though
only two of the three systems have a directly controlled partition. We take the
shared, substrate-independent law to be the U-shape with an interior optimum,
while the optimum's exact location reflects how much within-category variability
each world contains. We regard the shared
U-law, not a universal optimum, as the defensible cross-system claim: not common
variables, and certainly not common biology, but a common relationship between
how finely a system groups its past and how well it handles its future.
 
The two arms of the U-shape are not the same failure, and the framework names them
separately. The left arm (too few groups) is a violation of behavioural
sufficiency: groups grow until they merge experiences whose futures differ, and
the cost appears as false recall in the associative model and self-reactivity in
the immune model. The right arm (too many groups) is a violation of
generalization: groups become so narrow that a novel variant matches no address
well, so the system either responds from an inappropriate category or must build
a new one from scratch, and each group is in any case visited too rarely for its
future to be estimated from finite experience. The immune sweep
separates these directly, since protection against drifted variants and
self-reactivity are measured on the same repertoires and rise on opposite sides
of the optimum (Fig~\ref{fig:immune}B). Requirements~1 and~2 push in
opposite directions on granularity, and the interior optimum is where they
balance; requirements~3 and~4 shift the position of that optimum without
removing it.
 
\paragraph{Why the scaffold cannot be replaced by recomputation}

The four requirements characterize a useful grouping but do not explain why a
persistent scaffold is needed at all. If omitted structure could always be
reconstructed from later observations, memory would be only an efficiency
device. Whether recomputation can substitute for storage depends on the
experience process.
Let $\theta$ be a behaviourally relevant statistic, such as the distribution of
futures given a cue. We call $\theta$ \emph{recoverable} over horizon $T$ when
an estimator based on a single trajectory of length $T$ converges to it. This
yields the following trichotomy.

\begin{proposition}[Recoverability trichotomy]
\label{prop:recover}
Let $\tau_{\mathrm{mix}}$ denote the mixing time of the experience process and
$T$ the agent's lifetime.
\begin{enumerate}[leftmargin=1.6em,itemsep=1pt]
\item If the process is ergodic and $\tau_{\mathrm{mix}} \ll T$, then $\theta$
is recoverable: by the pointwise ergodic theorem the time average converges to
the ensemble average, and storage buys speed only.
\item If the process is ergodic but $\tau_{\mathrm{mix}} \gg T$, then $\theta$
is not recoverable within the lifetime: storage is mandatory for this agent,
though not in principle.
\item If the process supplies no convergent target for $\theta$, either
(3-a) because it drifts, so that the estimand itself moves on the timescale
over which it would have to be estimated, or (3-b) because time averages of
$\theta$ fail to self-average, remaining random variables in the limit of long
observation, then $\theta$ is not recoverable at any $T$: storage is the only
mechanism of retention.
\end{enumerate}
\end{proposition}
 
Case~(3) is stronger than ordinary decomposition into ergodic components. In a
component-decomposable process, time averages converge to the statistics of the
realized component, so continued observation can recover everything relevant to
that trajectory. Under weak ergodicity breaking, by contrast, time-averaged
observables remain random even at long times
\cite{Bouchaud1992,BelBarkai2005,HeBurov2008}; no unique target is available for
recomputation.
The distinction can be expressed computationally. \emph{Memoization} stores a
value that could be regenerated and purchases speed. \emph{Inscription}
records structure for which no convergent reconstruction is available and purchases possibility. Case~(1) is memoization; cases~(2) and~(3)
require inscription over the relevant lifetime. Biological memory does both, but only inscription
makes a persistent scaffold obligatory. 

The consequence for our framework is that consolidation is not optional
housekeeping. A system facing a drifting, non-self-averaging or slowly recurring
environment must commit structure to the scaffold at the time of encounter or lose it
permanently, and it must do so under the four requirements already stated,
because it cannot afford to inscribe everything and cannot recover what it
declines to inscribe. We note that memory necessity has a precedent of a different kind. In
computational learning theory, some concept classes provably cannot be learned
with bounded memory regardless of how many samples are drawn
\cite{Raz2018}. That obstruction is combinatorial, arising from the structure of
the hypothesis class under an ergodic (indeed independent) sample stream; ours
is measure-theoretic, arising from the structure of the process under an
unremarkable target. The two are complementary routes to the same conclusion,
that storage can be \emph{mandatory} rather than efficient.

\subsection*{Ablation study for the scaffold}
We tested the prediction directly following from Proposition \ref{prop:recover} in our ablation study with the scaffold. To simulate a lesion in a drifting environment, entries were deleted from a consolidated predictive
quotient (the scaffold of Fig~\ref{fig:willshaw}), held at bounded
capacity and rewritten from a recent-evidence window, so that a deleted entry is
genuinely absent. A matched sham store received the identical
episode stream without the lesion, which separates what the lesion destroyed
from what the world subsequently failed to supply. Ablating a third of the
addresses drove accuracy at those addresses from $0.947$ to the $0.502$ chance
floor. Under a stationary regime in which the ablated structure recurred on a
short interval, continuing experience restored it with no retraining and no
error signal: accuracy returned to $0.939$, regenerating $98\%$ of the deficit,
and the lesioned and sham stores became indistinguishable. When the same
structure was singular (written during warm-up and never encountered again,
in a process that remained stationary in aggregate), nothing came back
($\rho=-0.00$) while the sham store still held the content at $0.947$. The two
conditions differ in no respect except the recurrence time of the ablated
statistic, which is what Proposition~\ref{prop:recover} says should govern the
outcome.
 
\begin{figure}[!ht]
\centering
\includegraphics[width=\textwidth]{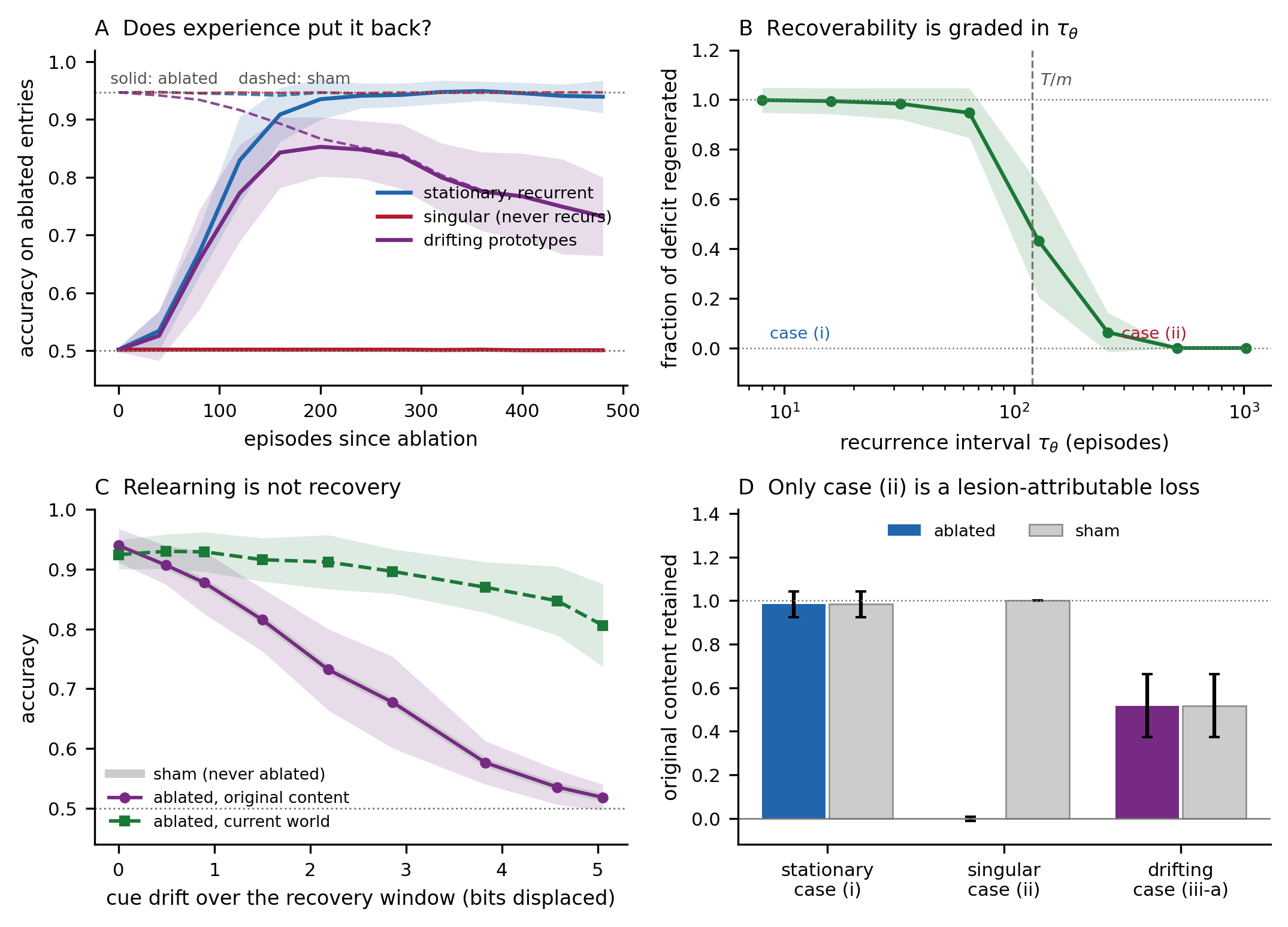}
\caption{{\bf Ablating scaffold entries: experience regenerates what it will
supply again, and only that.}
A third of the addresses of a consolidated predictive quotient are deleted at
$t=0$ and the store is then driven by continuing episodes; a sham store sees the
identical stream without the lesion. Accuracy is measured on held-out cues built
from the \emph{original} prototypes at the ablated addresses.
(A) Recovery trajectories. Stationary, short-recurrence structure returns to the
pre-lesion level; singular structure does not return at all, while its sham
retains it; drifting structure is rewritten quickly and then decays as the
target moves, the sham decaying with it. Solid, ablated; dashed, sham; bands
$\pm1$ s.d.
(B) Fraction of the lesion-induced deficit regenerated, against the recurrence
interval $\tau_{\theta}$ of the ablated statistic. The transition falls at the
horizon comparison $T/m$ (dashed), the point at which the remaining lifetime
stops supplying the evidence a rewrite needs.
(C) Under drift, accuracy on the original content declines identically for the
ablated and sham stores (the curves coincide), while competence on the current
world at the same addresses is preserved: relearning is not recovery.
(D) Ablated versus sham retention of the original content by regime. Only the
singular regime shows a lesion-attributable loss; in the drifting regime both
stores lose the content, which is what it means for the process to supply no
convergent target. Mean of $16$ stochastic runs.}
\label{fig:ablate}
\end{figure}

Recoverability was graded in that recurrence time rather than dichotomous
(Fig~\ref{fig:ablate}B). Sweeping $\tau_{\theta}$ over two decades moved
regeneration smoothly from complete to absent, with the half-recovery point at
$\tau_{\theta}=117$ episodes against the $T/m=120$ predicted by the evidence a
rewrite requires. The case~(1)/case~(2) boundary is a horizon comparison, and
the simulation puts it where the comparison says it should be. This is the
spectrum-over-queries reading of the trichotomy made concrete: one process, one
lifetime, and a recoverability verdict that depends on which statistic is asked
about.
Drift behaves like neither (Fig~\ref{fig:ablate}C,D). With cue prototypes
random-walking during the recovery window, the ablated entries were rewritten
promptly, and the lesioned and sham stores again became identical at every
drift rate tested the gap between them was numerically zero. Yet accuracy on the
content as originally encoded fell with drift for both, to $0.732$ at a mean
displacement of $2.2$ bits and to $0.519$, just above the chance floor, by $5$
bits, while competence on the \emph{current} world at those addresses stayed
high ($0.912$ and
$0.806$ respectively). The system was neither damaged nor idle; it had relearned
a different world. That dissociation is the operational signature separating
case~(3-a) from case~(2): under drift the loss is real but not
lesion-attributable, because the target that restoration would have to reach no
longer exists, whereas the singular case is a loss the lesion alone caused and
that no quantity of further experience repairs. It also shows why relearning
cannot be read as recovery, and why an experiment that measures only post-lesion
performance on the current environment will mistake the second regime for the
first.

% ============================================================
\section*{Discussion}

%\paragraph{When is a memory worth keeping?}
%Storage can also be mandatory for combinatorial reasons: some concept classes cannot be learned with bounded memory even under independent sampling \cite{Raz2018}. That obstruction concerns the hypothesis class, whereas Proposition~\ref{prop:recover} concerns the generating process. In either case, a retained memory must still justify its cost. Over a future horizon, its value is the error it prevents minus the cost of writing and maintaining the scaffold.
%The formal analysis in \nameref{S1_Appendix} decomposes the shortfall from this ideal value into five penalties: limited model flexibility, finite experience, slow--fast approximation, leakage, and environmental drift.

\paragraph{What should be forgotten.}
Now we return to the question in our title. The four requirements determine
which \emph{distinctions} should be preserved: experiences may be merged when
they induce the same distribution over futures, whereas distinctions that do
not affect what happens next need not occupy persistent structure
\cite{shalizi2001computational}. Proposition~\ref{prop:recover} supplies a
complementary criterion for deciding which \emph{records} may be discarded:
forget what ongoing experience can reconstruct, and retain what it cannot.
These criteria are independent. The location of a reliably replenished food
source may remain behaviorally important yet be recoverable within days,
whereas a single near-fatal encounter may support only a coarse distinction
but remain irreplaceable because the experience needed to relearn it is
unlikely to recur within a lifetime. Forgetting is not necessarily a failure
of maintenance; it can be an adaptive allocation of limited memory resources
toward behaviorally valuable structure
\cite{richards2017persistence,nairne2010adaptive}.

Our account yields a testable dissociation from frequency-based theories of
memory maintenance \cite{anderson1991reflections}. Partition consolidated memories according to whether their
content can be reconstructed from ongoing experience: rapidly recurring
regularities fall under case~(1), whereas singular, drifting, or slowly
recurring structures fall under cases~(2)-(3). After targeted disruption,
memories in case~(1) should re-emerge without explicit retraining, while those
in cases~(2)-(3) should remain lost. Maintenance investment and decay should
track non-recoverability rather than retrieval frequency: a rarely
retrieved but non-regenerable memory should be protected more strongly than a
frequently retrieved but readily recoverable one. A frequency-based account
predicts the reverse ordering, while a purely efficiency-based account predicts
no such dissociation. The simulations reported in Fig. \ref{fig:ablate} test this prediction by
ablating scaffold entries under stationary and nonstationary generative
regimes; an analogous biological test would compare recovery after disrupting
memories for statistically regular versus singular episodes. What we should
forget is governed by two estimable quantities: whether a distinction
changes the future, and whether experience will supply it again.

\paragraph{Relation to predictive-state and biological memory accounts.}
Causal states, $\epsilon$-transducers, predictive rate-distortion, and
probabilistic bisimulation provide abstract notions of predictive compression
and controlled-state equivalence
\cite{shalizi2001computational,barnett2015computational,
still2010optimal,givan2003equivalence}. We use these constructions as
antecedents and ask how an approximate predictive quotient can be realized in
slowly changing biological structure. In the present simulations, inputs are
assigned exogenously and independently of the latent disturbances by
construction, so conditioning on an input agrees with intervention on that
input within the simulator \cite{pearl2009causality}. Behavioral data generated
under a policy would instead require randomized input assignment, or assumptions
that identify the intervention from observational trajectories together with
appropriate off-policy correction \cite{precup2000eligibility}. Off-policy
correction alone would not resolve hidden confounding.

Complementary learning systems theory \cite{McClelland1995} separates fast
episodic from slow semantic learning; scaffold-flow adds a criterion for what the
slow system should retain: the coarsest grouping that remains sufficient,
generalizable, non-leaky, and affordable. Attractor models
\cite{Hopfield1982,Amit1989} represent memories as stable states; our spectral
formulation instead treats them as metastable categories whose barriers can
change during consolidation, reconsolidation, and forgetting. Immune repertoire
models \cite{Perelson1997,DeBoer2013} describe affinity and abundance dynamics;
the grouping view supplies a normative account of cross-reactive breadth.

\paragraph{Generalization under inscription.}
Requirement~2 asks that a grouping support experiences not encountered during
its construction. Proposition~\ref{prop:recover} shows that this demand takes
dual forms in the two regimes. Memoization is the setting assumed by classical
learning theory \cite{geman1992neural}: many effectively independent samples are available for a fixed
query $\theta$, empirical estimates concentrate, and generalization is measured
as excess population risk. Inscription reverses the scarcity. There is one
sample, one life, and an open-ended set of future queries. The scarce resource
is not sample size but coverage of queries not yet posed. A quotient
written from the experiences that happened to construct it must remain useful
for queries on which it was never tested. The relevant requirement is 
uniform recoverability over a query class $\Theta$, not pointwise recoverability
for a single $\theta$. Even when every $\theta\in\Theta$ is individually
recoverable, non-uniform rates can leave some queries unresolved at every finite
horizon $T$.
Predictive sufficiency controls this gap directly. If
$P(\text{future}\mid h)=P(\text{future}\mid q(h))$, then
$\mathbb E[f\mid h]=\mathbb E[f\mid q(h)]$ for every bounded future-measurable
functional $f$. More generally, if the conditional future laws differ by at
most $\delta$ in total variation within each quotient cell, then
$\bigl|
\mathbb E[f\mid h]-\mathbb E[f\mid q(h)]
\bigr|
\le
2\|f\|_\infty\delta$
uniformly over bounded queries. Query generalization is governed by
the predictive-sufficiency defect together with the lumpability defect measured
by leakage. Under case~(3), the same statement holds quenched, with the
realized environment's conditional law as the target. The design implication is
that consolidation should minimize divergence between future laws rather than
error on any single reconstruction target: a quotient optimized for one query
may fail arbitrarily on another, whereas a quotient preserving the future law
serves all measurable queries derived from it.

Inscription nevertheless lacks the consistency guarantee available to
memoization. A memoizing system can wait for concentration, allowing premature
commitments to wash out. An inscribing system must commit at encounter time from
a trajectory that may never become representative, so generalization depends
primarily on the inductive bias of the write policy \cite{griffiths2010probabilistic}. Three consequences follow.
First, the two sides of the granularity trade-off are asymmetric. A coarse
partition risks an irreversible false merge that deletes distinctions between
different futures, whereas an overly fine partition primarily incurs capacity
and coverage costs. This asymmetry motivates hysteretic write control: merge
only under strong evidence and reverse the decision only after a distinct lower
threshold is crossed. Second, prior structure is not optional; it substitutes
for samples that will never be observed. As more of the environment falls under
cases~(2)-(3), competence must rely increasingly on write-time priors rather
than accumulated statistics. Third, persistence should be allocated according
to non-recoverability rather than frequency. Under a capacity constraint, the
system should preferentially inscribe queries with
$\tau_\theta\gtrsim T$, so that a rare but non-recoverable event can outrank a
frequent self-averaging one. Frequency-weighted storage is optimal only in the
memoization regime, where storage primarily buys amortization. This yields a
direct, falsifiable contrast with replay-frequency and reconstruction-based
accounts, both of which privilege frequent experiences by construction.

\paragraph{Experimental predictions with matched controls.}
In neural systems, consolidation should move category-associated modes toward
unity, strengthen within-category contraction, and reduce episode-specific
information while preserving behavioural consequence. A matched manipulation
that changes gain without changing category structure should alter contraction
but not slow-mode count. In immune systems, memory-compartment breadth should
track pathogen variability; imprinting should appear as a grouping that no
longer updates; and memory and effector compartments may realize category
identity and cached response at different levels of the same hierarchy. Across
both systems, manipulating the number or variability of distinct futures should
shift the interior optimum in Fig.~\ref{fig:cross}.

\paragraph{Limitations.}
The simulations test quantitative consequences of the proposed architecture but
are not biophysically complete models. The immune simulator includes clonal
selection, receptor mutation and breadth, and tolerance, but omits spatial
germinal-centre organization, detailed T-cell help, isotype, and compartment
structure. Its self-reactivity result locates a predicted trade-off rather than
establishing a mechanism of autoimmunity. The immune optimum above $r^*$ may
reflect within-family antigenic variability, which requires independent
measurement of the effective number of immune futures. The slow--fast reduction
assumes timescale separation and may become set-valued under strong
metastability. Leakage depends on a chosen metric between future distributions,
whose biological calibration remains open. Finally, our experiments are passive
or exogenously driven; separating prediction from intervention will require
randomized inputs or causal abstractions.

\paragraph{Conclusion.}
Memory consolidation is the decision of which distinctions in a transient
experience should be written into structure that outlives it. We have argued
that the selected grouping must preserve consequential differences, cover novel
variants, support autonomous coarse dynamics, and justify its physical cost.
Across an associative memory, a metastable neural network, and a simulated
immune repertoire, these constraints produced the same qualitative law: future
risk was minimized at an intermediate granularity. Neural and immune memory
share neither molecules nor detailed equations; they share a computational
problem whose solution is a predictive, dynamically usable, and affordable
coarse-graining of the past.

% ============================================================
\section*{Materials and methods}
\label{sec:mm}
 
\subsection*{Formal statement of the architecture}
Fast state $X_t$ and slow scaffold $S_t$ evolve as
$dX_t=f(X_t,S_t,U_t)\,dt+\sigma(X_t,S_t,U_t)\,dW_t$ and
$dS_t=\varepsilon G(S_t,X_t,U_t,E_t)\,dt+\sqrt{\varepsilon}\,\Sigma_S\,dB_t$,
with $0<\varepsilon\ll1$, punctuated by consolidation events
$S^{+}=\mathcal Q(S^{-},\Gamma)$. The scaffold determines the generator and
future path law, $S\mapsto\mathcal L_S\mapsto K^T_S$. Histories satisfy exact
finite-horizon predictive equivalence, $h\sim_T h'$, when
$P^T(\cdot\mid h,u)=P^T(\cdot\mid h',u)$ for every relevant future input $u$.
Empirical clusters use nonzero tolerances and need not form a transitive exact
quotient. Leakage is
$\Lambda_T(\pi)=\operatorname{ess\,sup}_{\pi(x)=\pi(x')}
d_{\mathsf P}(\pi_\#K_x^T,\pi_\#K_{x'}^T)$; it vanishes exactly under
lumpability. %Formal definitions and proofs appear in \nameref{S1_Appendix}.

\subsection*{Consolidation algorithm}
Algorithm~\ref{alg:consolidate} implements the online procedure described in the
Results. Each group stores an address $a_g$, response program $r_g$, and future
statistics $F_g$. Split and merge tests act on $F_g$, address updates extend or
contract coverage of novel cues, and scaffold updates reduce
$\Lambda_T(\pi)$. By Eq.~\eqref{eq:excess}, this last step controls the excess
error induced by acting on group labels rather than microscopic states.

\begin{algorithm}[!ht]
\caption{Scaffold--flow consolidation}
\label{alg:consolidate}
\begin{algorithmic}[1]
\State \textbf{maintain} groups $\{g\}$, each with address $a_g$, response $r_g$,
and observed future statistics $F_g$
\For{each new experience $h$ with observed future $\xi$}
  \State $g \gets$ group whose address best matches $h$ \Comment{recall}
  \State update $F_g$ with $(h,\xi)$
  \State \textbf{if} no group matches \textbf{then} create a new group
  \Statex
  \State \textbf{// Requirement 1: do not merge different futures}
  \If{spread of futures within $g$ exceeds $\eta_{\text{split}}$}
    \State split $g$ into subgroups with internally consistent futures
  \EndIf
  \Statex
  \State \textbf{// Requirement 2: make the address cover novel variants}
  \If{$\xi$ is consistent with $F_g$}
    \State broaden $a_g$ towards $h$ at rate $\eta_{\text{gen}}$
    \Comment{cover unseen cues sharing this future}
  \Else
    \State contract $a_g$ away from $h$ \Comment{exclude variants with other futures}
  \EndIf
  \Statex
  \State \textbf{// Requirement 3: keep the abstraction non-leaky}
  \State estimate leakage $\Lambda$ (Eq.~\eqref{eq:leak}); adjust the scaffold to
  reduce it
  \Statex
  \State \textbf{// Requirement 4: do not pay for redundant groups}
  \If{two groups have future statistics closer than $\eta_{\text{merge}}$}
    \State merge them
  \EndIf
  \State write $\{a_g, r_g\}$ into the physical scaffold; decay unused groups
\EndFor
\end{algorithmic}
\end{algorithm}
 
The thresholds $\eta_{\text{split}}$ and $\eta_{\text{merge}}$, the
address-broadening rate $\eta_{\text{gen}}$, and the decay
rate for unused groups are instantiated separately in each substrate, as
described in the simulation subsections below. The broadening step has a
concrete reading in each: widening the basin of an attractor so that degraded
cues still fall into it, and raising the cross-reactive breadth of a selected
clone so that drifted antigens are still bound.
 
\subsection*{Associative memory simulations}
Binary scaffolds of size $192\times192$; patterns with $6$ active units; six
response categories $\times$ four cue clusters; experiences generated by flipping
each pattern bit with probability $0.22$; recall threshold $0.7$ of the cue
weight. Error was $(w\cdot\text{false alarm}+\text{miss})/(w+1)$ on held-out
corruptions, with $w\in\{1,5\}$. Split/merge agglomerated cue clusters whose
observed futures coincided. Six runs per condition.
 
\subsection*{Modern Hopfield simulations}
Cues were unit vectors in $d=128$ dimensions drawn around $24$ prototypes ($6$
predictive categories $\times$ $4$ cue clusters) by adding Gaussian noise of
standard deviation $0.09$ per component before renormalization. Responses were
sparse binary vectors of length $24$ with $6$ active bits per category, observed
with each bit flipped independently with probability $0.30$. Retrieval was the
modern Hopfield update in attention form,
$\hat v = V^{\top}\mathrm{softmax}(\beta K q)$, with $\beta=12$ except in the
temperature sweep. Risk was $(w\cdot\text{false alarm}+\text{miss})/(w+1)$ at
threshold $0.5$ on held-out cue corruptions ($8$ per cue cluster), with $w=1$
unless stated. The four strategies differ only in $(K,V)$: episodic storage uses
every cue with its own observed response; the metric quotient uses $r$ cue-space
$k$-means centroids paired with the mean response of their members; the predictive
quotient uses the $24$ fine cue centroids as keys and, as values, the mean response
pooled over each predictive group; split/merge builds the same structure but
agglomerates fine clusters whose \emph{thresholded} mean responses agree to within
$0.20$ per bit. Thresholding matters: comparing raw means fails at high load, where
they approach the flip rate and the between-category gap falls below any fixed
tolerance. Eight seeds per condition; the granularity sweep used $3$ episodes per
cue cluster and the temperature sweep $16$.
 
\subsection*{Neural simulations}
A recurrent rate network of $N=500$ units stored $K=6$ random sparse patterns
$\xi^k\in\{0,1\}^N$ with coding level $f=0.12$, through the covariance rule
$J=\tilde\xi^{\top}\tilde\xi/[Nf(1-f)]$, $\tilde\xi=\xi-f$, with the diagonal
removed. Dynamics were
$\dot x = -x + gJ\phi(x) - w_I(\langle\phi\rangle-f) + \xi_{\text{noise}}$,
with $\phi(x)=[1+e^{-\beta x}]^{-1}$, $\beta=8$, $w_I=20$, integrated by
Euler--Maruyama at $\mathrm dt=0.1$ and recorded every $10$ steps.
Consolidation was modelled as an increase in the recurrent gain $g$, which
deepens the basins without altering what is stored.
 
Because the escape barrier of an attractor network grows with $N$, independent
per-unit noise makes spontaneous transitions unobservable at this size, and
forcing itinerancy with firing-rate adaptation instead would clock every escape
by the adaptation time constant rather than by the noise - the wrong model for
a Kramers test. We used structured (shared) variability,
$\xi_{\text{noise}} = \sigma_i\eta(t) + \sigma_s\tilde\xi^{\top}\eta_K(t)
/\sqrt{Nf(1-f)}$ with $\sigma_i=0.05$, so that fluctuations of the collective
coordinate do not vanish as $1/\sqrt N$; $\sigma_s$ is the noise amplitude swept
in Fig.~\ref{fig:neural}C.
 
All spectral and kinetic quantities were estimated from the simulated firing
rates of $250$ randomly chosen units, with no reference to the stored patterns.
Activity vectors were clustered into $250$ microstates by $k$-means; the
transition operator was the row-normalized count matrix at lag $5$ (four
independent trajectories of $8\times10^5$ steps, pooled); macrostates were
obtained by $k$-means on its leading right eigenvectors (PCCA-style); implied
timescales are $-\tau/\log\lambda_i$. Mean first-passage times to a macrostate
$B$ were obtained by solving $(\mathrm I-T_A)t=\mathbf 1$ on the complement and
averaging over the source macrostate under the stationary distribution, and were
compared with first-passage times measured directly from the discrete state
sequence at the same temporal resolution. Stored patterns were used only
\emph{post hoc}, to report the purity of the recovered macrostates.
 
\subsection*{Hierarchical landscape}
A periodic collective coordinate discretized into $120$ microstates carried the
nested potential $V(x)=-B_c\cos(2\pi kx)-B_f\cos(2\pi kmx)$ with $k=3$, $m=4$,
$B_c=0.8$, $B_f=0.4$ and temperature $\theta=0.35$, evolving by nearest-neighbour
Metropolis moves, which are reversible with respect to $e^{-V/\theta}$. Implied
timescales were $-\tau/\log\lambda_i$ from the exact transition matrix. For the
prediction task, macrostates were $r$ contiguous arcs offset by half a well width
so that their boundaries fall on barriers rather than inside wells, and the risk of
granularity $r$ at horizon $\tau$ was the out-of-sample negative log-likelihood of
the microstate occupied $\tau$ steps later under
$p(j\mid i)=\sum_b \hat T_r[a(i),b]\,w(j\mid b)$, where $a(i)$ is the macrostate of
$i$, $\hat T_r$ the macrostate transition matrix estimated at lag $\tau$, and
$w(j\mid b)$ the stationary distribution restricted to macrostate $b$. Every
$(r,\tau)$ pair was estimated from the same budget of $n=3000$ transition pairs
held out from a trajectory of $7\times10^5$ steps, and averaged over $12$ seeds, so
that coarse and fine models compete on equal evidence rather than on sample size.
The bounded-capacity term added to the likelihood was the description length
$\tfrac{r(r-1)}{2}\cdot\tfrac{\log n}{n}$ of the free transition parameters.
Granularities tested were $r\in\{1,2,3,4,6,8,10,12,15,20,24,30,40,60\}$ and
horizons $\tau$ from $1$ to $5\times10^4$. The recursion test compared implied
timescales of the coarse-grained operator at $r=12$ (lag $20$) and $r=3$
(lag $200$) with those of the microscopic operator.
 
\subsection*{Immune simulations}
Stochastic germinal-centre affinity maturation in a two-dimensional antigen shape
space. Each B cell carried a receptor position $r$ and breadth $w$, binding
antigen $x$ with affinity $(w_{\mathrm{ref}}/w)^{\alpha}\exp(-\|r-x\|^2/2w^2)$,
$w_{\mathrm{ref}}=0.30$, $\alpha=1$, so breadth trades against peak affinity. Six
antigenically variable pathogen families ($r^*=6$) each presented five strains;
drifted variants and a disjoint set of self antigens were held out. A naive
repertoire of $2500$ cells underwent central tolerance (deletion of strongly
self-reactive clones), then one germinal centre per family ran eight rounds of
proliferation, somatic hypermutation of both $r$ and $w$, peripheral tolerance,
and affinity-based selection retaining the top fraction. A single parameter
$\lambda\in[0,1]$ set the selection criterion from best single-strain affinity
($\lambda=0$) to worst-case affinity across strains ($\lambda=1$); receptor
breadth was not imposed but evolved under this pressure. Emergent granularity was
the number of memory-receptor clusters at a resolution set by their own breadth.
Total risk was $(1-\text{variant protection})+3\times\text{self reactivity}
+0.01\times\text{groups}$. Twelve stochastic runs per condition. %Full code is in \nameref{S2_Code}.
 
\subsection*{Cross-system comparison}
Group count $r$ was swept in each system and normalized by $r^*=6$. Below $r^*$,
categories with distinct futures were merged; above $r^*$, categories were split
so that each was estimated from fewer experiences, with an explicit per-category
maintenance cost ($0.020$ per group). The associative curve is the value-side
granularity sweep of Fig.~\ref{fig:hopfield}C with that maintenance cost added. For the recurrent network the state space
was the same $k$-means partition of recorded population activity used in
Fig.~\ref{fig:neural} ($120$ cells), grouped at granularity $r$ by clustering the
cell centroids; the score was the out-of-sample predictive log-loss of the
microstate a lag $\tau=25$ later given the current group, averaged over $10$
resamplings of $2500$ pairs. Curves were min-max normalized within system for
display only; optima were computed on unnormalized values.
 
\subsection*{Code availability}

{\bf Simulation suite.} Python implementation of all experiments, including an
annotated Jupyter notebook that builds the stochastic affinity-maturation
simulator step by step (binding trade-off, germinal-centre reaction, tolerance,
emergent-granularity readout, and recall dynamics) are available from the supporting information accompanying this submission.

\section*{Competing interests}
The author has declared that no competing interests exist.
 
\section*{Author contributions}
Conceptualization, methodology, software, formal analysis, investigation,
visualization, and writing: Xin Li.

\section*{Declaration of AI Usage}
Large language models (Claude Opus 5 by Anthropic and ChatGPT 5.6 by OpenAI) were used as assistive tools during the preparation of this work in the following capacities:
(1)~drafting and iterating on experimental code, including Jupyter notebooks for benchmark implementations and baseline adaptations;
(2)~generating initial drafts of tables, and figure captions, which were subsequently reviewed, revised, and verified by the authors;
(3)~debugging code, checking numerical consistency between notebook outputs and manuscript tables, and suggesting structural organization for the experimental sections.
All theoretical contributions (definitions, theorems, proofs), experimental design decisions, scientific interpretations, and final manuscript content were conceived, validated, and approved by the authors.
The authors take full responsibility for the correctness and integrity of the published work.

% ============================================================

\end{document}